\documentclass[12pt]{spieman}  
\usepackage{amsmath,amsfonts,amssymb}
\usepackage{graphicx}
\usepackage{setspace}
\usepackage{tocloft}
\usepackage{amsmath}

\title{Artifact-less Coded Aperture Imaging in the X-ray Band with Multiple Different Random Patterns}

\author[a,*]{Tomoaki Kasuga}
\author[a,b,c]{Hirokazu Odaka}
\author[a]{Kosuke Hatauchi}
\author[a]{Satoshi Takashima}
\author[a]{Tsubasa Tamba}
\author[a]{Yuki Aizawa}
\author[a,b]{Soichiro Hashiba}
\author[a,b]{Aya Bamba}
\author[d,e]{Yuanhui Zhou}
\author[d,e,f]{Toru Tamagawa}
\affil[a]{Department of Physics, Graduate School of Science, The University of Tokyo, 7-3-1 Hongo, Bunkyo-ku, Tokyo 113-0033, Japan}
\affil[b]{Research Center for the Early Universe, School of Science, The University of Tokyo, 7-3-1 Hongo, Bunkyo-ku, Tokyo 113-0033, Japan}
\affil[c]{Kavli IPMU (WPI), UTIAS, The University of Tokyo, 5-1-5 Kashiwanoha, Kashiwa, Chiba 277-8583, Japan}
\affil[d]{RIKEN Cluster for Pioneering Research, 2-1 Hirosawa, Wako, Saitama 351-0198, Japan}
\affil[e]{Department of Physics, Tokyo University of Science, 1-3 Kagurazaka, Shinjuku-ku, Tokyo 162-8601, Japan}
\affil[f]{RIKEN Nishina Center, 2-1 Hirosawa, Wako, Saitama 351-0198, Japan}

\cftpagenumbersoff{figure}
\cftpagenumbersoff{table} 
\begin{document} 
\maketitle

\begin{abstract}
The coded aperture imaging technique is a useful method of X-ray imaging in observational astrophysics. However, the presence of imaging noise or so-called artifacts in a decoded image is a drawback of this method. We propose a new coded aperture imaging method using multiple different random patterns for significantly reducing the image artifacts. This aperture mask contains multiple different patterns each of which generates a different artifact distribution in its decoded image. By summing all decoded images of the different patterns, the artifact distributions are cancelled out, and we obtain a remarkably accurate image. We demonstrate this concept with imaging experiments of a monochromatic 16~keV hard X-ray beam at the synchrotron photon facility SPring-8, using the combination of a CMOS image sensor and an aperture mask that has four different random patterns composed of holes with a diameter of 27 $\mu$m and a separation of 39 $\mu$m. The entire imaging system is installed in a 25 cm-long compact size, and achieves an angular resolution of $< 30^{\prime \prime}$ (full width at half maximum). In addition, we show by Monte Carlo simulation that the artifacts can be reduced more effectively if the number of different patterns increases to 8 or 16.
\end{abstract}


{\noindent \footnotesize\textbf{*}Tomoaki Kasuga,  \linkable{tomoaki.kasuga@phys.s.u-tokyo.ac.jp} }

\begin{spacing}{2}   

\section{Introduction}
\label{sect:intro}  
The coded aperture imaging \cite{Ables68, Dicke68, Caroli87} is one of the useful methods for imaging X-rays. This method is employed by the {\it Swift}-BAT \cite{Gehrels04, Barthelmy05} and  the {\it INTEGRAL}-IBIS \cite{Winkler03, Ubertini03, Goldwurm03} and -SPI \cite{Winkler03, Vedrenne03} for astronomical observatories, and gamma-ray imaging systems on ground for nuclear medicine and nondestructive inspection. In the imaging process, we obtain the projection of an aperture pattern by a position-sensitive detector, and estimate the directional distribution of photon sources based on the projected pattern. In contrast to focused imaging using a multilayer coated mirror such as the {\it NuSTAR} \cite{Harrison13} and the {\it Hitomi}-HXI \cite{Nakazawa18}, the coded aperture does not need a complicated optics and a long focal length. This simplicity allows us to build a compact imaging system that can be installed in an ultra-compact satellite like a CubeSat. \par
There are two major categories of the coded aperture patterns. One is a category of patterns that have a nature with the characteristics of uniform redundancy. This is an application of the difference set and the pseudo noise generation \cite{Baumert69, Caroli87} to computational imaging. Several patterns have been proposed historically: e.g., M-sequence \cite{MacWilliams76}, Uniformly Redundant Arrays (URA) \cite{Calabro68, Fenimore78a}, and Modified URA (MURA) \cite{Calabro68, Gottesman89}. Since these patterns make use of a property that the autocorrelation is the delta function, we should use periodic repetition of these patterns for realizing a modulo operation in the convolution operation in an image decoding process. Then the reconstructed image in the fully coded field-of-view (FC-FoV) of this category of the patterns is completely artifact-less except for statistical fluctuations \cite{Badiali85, Caroli87}. On the other hand, concerning the outside of the FC-FoV, which is named partially coded FoV (PC-FoV), we have eight strong artifacts and weaker but structural ones in the PC-FoV due to the periodicity, when a photon source is in the FC-FoV. For example, the artifacts have at most half intensity of the source peak in the IBIS detector on {\it INTEGRAL}, which employs the MURA 53 $\times$ 53 \cite{Goldwurm03}. A worse problem is that when the source is located in the PC-FoV, confusing artifacts appear even in the FC-FoV \cite{Badiali85}. Such artifacts induce serious uncertainties in the reconstructed image. \par
An alternative way is using a category of the random patterns. This pattern is coded by a two-dimensional array of random numbers sampled from \{0, 1\}, and we can straightforwardly obtain higher randomness with a larger number of elements. In the pattern, artifacts in the decoded image are generally not structural and are not remarkable. This is a clear advantage of the random patterns, and it is apparent in images obtained with the BAT detector on {\it Swift}, which adopts a random pattern composed of $\sim$52,000 lead tiles \cite{Gehrels04, Barthelmy05}. However, artifacts can be distributed in the entire FoV including the FC-FoV, in which an on-axis primary target source may come for pointing observation \cite{Caroli87}. This weak point of the random patterns becomes significant due to insufficient randomness when the number of the coded aperture elements is small. Particularly for a narrow FoV configuration, the elements number should be limited and thus it is important to gain high randomness of the pattern under the restriction of the image size. \par
The angular resolution is another important factor of the coded aperture imaging. Since existing missions such as {\it Swift} have been aimed at the all-sky survey of hard X-ray sources, their instruments have large FoVs by the sacrifice of the imaging resolution. They are typically 10 times or more inferior to the focused imaging method like {\it NuSTAR}, whose angular resolution is $18^{\prime \prime}$ (FWHM) \cite{Harrison13}. The angular resolution of the coded aperture imaging, in principle, can be improved by adopting even finer element sizes of the coded apertures and the detector pixels. \par
In this paper, we propose a new method to reduce artifacts that appear in images decoded with the random pattern coded aperture. Section \ref{sec:concept} describes the concept of our method , which simultaneously uses multiple images with different random patterns. We demonstrate this concept by hard X-ray beam experiments using our compact-scale imaging system for narrow FoV observations, as described in Section \ref{sec:exp}. Section \ref{sec:results} shows the results of these experiments. Here, we also demonstrate that our coded aperture system realizes a fine imaging resolution comparable to {\it NuSTAR} even in dimensional limitations of a compact-scale satellite of 25 cm. In Section \ref{sec:discuss}, we also conduct Monte Carlo simulations for further discussions and discuss practical issues in observations. We give our conclusions in Section \ref{sec:concl}.

\section{Coded Aperture Imaging with Multiple Different Random Patterns}
\label{sec:concept}
Mathematically, the coded aperture imaging is described as a convolution operation of a sky image $S$ with an aperture pattern $A$. In the encoding process corresponding to the measurement, the detected image $D$ is written by: 
\begin{equation}
D = A * S + B \ ,
\label{eq:encoding}
\end{equation}
where $*$ denotes the convolution operator and $B$ is the background on the detector. The purpose of the imaging is to estimate $S$ by computing a decoded image $\tilde{S}$. This decoding process is described with a decoding pattern $\tilde{A}$, which can be generated from $A$, as:
\begin{eqnarray}
\begin{cases}
\tilde{S} = \tilde{A} * D \\
\tilde{S}_{(i-k,j-l)} = (\tilde{A} * D)_{(i-k,j-l)} = \sum_{(i,j)} \sum_{(k,l)} \tilde{A}_{(i,j)} D_{(k,l)} 
\end{cases} . 
\label{eq:convolution}
\end{eqnarray}
Substituting Equation \ref{eq:encoding} for Equation \ref{eq:convolution}, $\tilde{S}$ is represented as:
\begin{equation}
\tilde{S} = (\tilde{A} * A) * S + \tilde{A} * B \ .
\label{eq:decoding}
\end{equation}
Ignoring $B$, which is independent of $A$, $\tilde{S}$ is reconstructed into the original image $S$ if the convolution $\tilde{A} * A$ is approximately the $\delta$ function. However, the difference between $S$ and $\tilde{S}$ results in artifacts. Since artifacts are originated from and unique to the pattern structure $A$, they are regarded as systematic errors of the coded aperture imaging. Such artifacts cause over- and under- estimations of the source intensity $S$. In the case of random patterns, $\tilde{A} * A$ converges to the delta function with the degree of randomness. Thus, the problem is reduced to how to increase the number of the pattern elements. 
\begin{figure}[H]
	\centering
	\includegraphics[width = 8 cm]{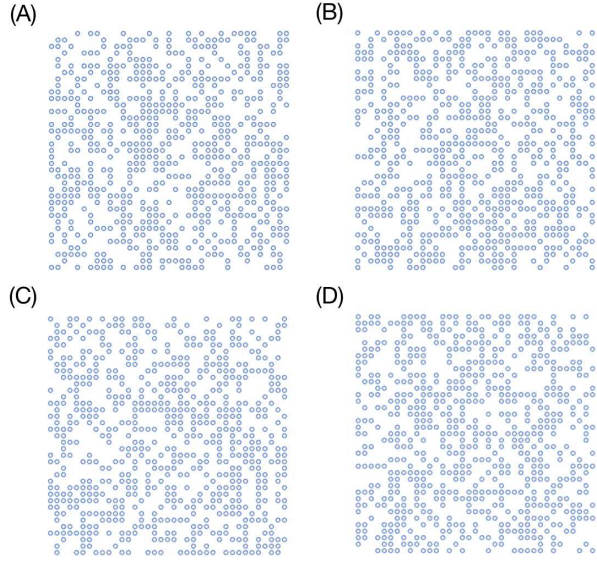}
	\caption[]{Example of multiple different random patterns. They are different only in terms of the configuration of apertures. This set of patterns was used for the experiments described in Section \ref{sec:exp}.}
	\label{patterns}
\end{figure}\par
We consider a coded aperture imaging system with a high-resolution and narrow-FoV configuration. The diameter of each aperture and the total pattern size should be determined by scientific requirements for the angular resolution and the FoV size. Since a general narrow FoV configuration does not need a large size of the entire system unit compared to the total size of its satellite, we can gain the effective area by configuring them parallelly. 
Random patterns have $2^{N^2}$ cases of varieties when the pattern size is $N \times N$. Figure \ref{patterns} shows one example. These four patterns are made randomly and independently, but the pattern size of each is equally set to be $37 \times 37$. The number of the aperture elements is 684 in order to keep the aperture ratio almost 50~\%. Here, we propose a simple idea to use this $2^{N^2}$ variation. Configuring parallelly some different random patterns with the same size instead of exactly same patterns, they give independent decoded images $\tilde{S}_p$ with the same FoV size and the angular resolution, where $p$ is a label for a pattern. They can be addable and a new summed image $\tilde{S}$ is given by:
\begin{equation}
\tilde{S} = \sum_{p} \tilde{S}_p \ .
\end{equation}
Due to the different artifact distribution in each $\tilde{S}_p$, their contribution should be canceled out in the summed $\tilde{S}$. It is important that the decoding is processed independently for each decoding pattern $\tilde{A}$, and is simply implemented in parallel computing.

\section{Experiments}
\label{sec:exp}
\begin{figure}[H]
	\centering
	\includegraphics[width = 8 cm]{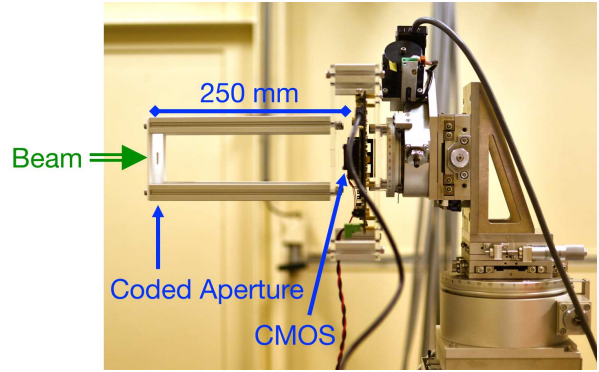}
	\caption[]{The setup for X-ray beam experiments. }
	\label{setup}
\end{figure}
In order to demonstrate the concept of the coded aperture imaging with multiple different random patterns, we performed X-ray beam experiments. Figure \ref{setup} shows the experimental setup. We configured the coded aperture plane in front of a CMOS imager with a distance of 250~mm. Changing the relative direction of the beam to the detector plane, we took three datasets for different imaging types of photon sources, as summarized in Table \ref{datasets}.
\begin{table}[ht]
\caption{Datasets in our experiments. The detected event counts are of unit dataset per each pattern before summing processes described in Section \ref{sec:results}.} 
\label{datasets}
\begin{center}       
\begin{tabular}{|c|c|c|} 
\hline
\rule[-1ex]{0pt}{3.5ex}  & Photon Source & Detected Event Counts\\
\hline\hline
\rule[-1ex]{0pt}{3.5ex}  (i) & a single point-like source on-axis & $\sim$13,000 \\
\hline
\rule[-1ex]{0pt}{3.5ex}  (ii) & two point-like sources on-axis \& $105^{\prime \prime}$ apart & $\sim$25,000 \\
\hline
\rule[-1ex]{0pt}{3.5ex}  (iii) & a circular source with a radius of $60^{\prime \prime}$ & $\sim$250,000 \\
\hline
\end{tabular}
\end{center}
\end{table} 

We made a coded aperture mask with the four different random patterns described in the previous section and shown in Figure \ref{patterns}. This mask is made of a 100~$\mu$m-thick SUS304 board and the apertures are created as 27~$\mu$m-diameter holes on it. To keep the mechanical strength, the pitch of the apertures is set to be 39~$\mu$m. Therefore, the entire pattern size of each pattern is $\sim$1.5~mm-squared. We arranged these four patterns parallelly on the SUS304 board with a margin of 500~$\mu$m in a $2 \times 2$ configuration. Then the total size of the multiple random patterns is $\sim$3.5~mm. We also define a detector area  in which the encoding image is obtained for each pattern as a direct projection to the detector plane along the plane normal with the same geometric size. It should be noted that this aperture mask is for the concept demonstration in this work, not for practical astrophysics observations. \par
We use the balanced correlation method \cite{Fenimore78a} as a decoding method. As the geometry of the apertures does not affect the quality of imaging \cite{Fenimore81, Sims84}, we use the ``delta decoding method'' \cite{Fenimore81}, where we assume that each photon comes through the center of the aperture circle. In the balanced correlation method, $\tilde{A}_{(i,j)}$ follows
\begin{eqnarray}
\tilde{A}_{(i,j)} = 
\begin{cases}
1 \ (A_{(i,j)} \ \mathrm{is~an~aperture.}) \\
-1 \ (A_{(i,j)} \ \mathrm{is~a~mask.})
\end{cases} ,
\end{eqnarray}
for an aperture fraction of 50~\% \cite{Fenimore78b}, where $(i,j)$ denotes indices of an aperture element. We also use the detected value $D_{(k,l)}$ at a detector element $(k,l)$, where the background $B_{(k,l)}$ is already subtracted. Using the central coordinates $(a_{x(i)}, a_{y(j)})$ of the coded aperture element $(i,j)$ and $(d_{x(k)}, d_{y(l)})$ of the detector element $(k,l)$, the photon direction $(s_x,s_y)$ which reaches the detector element $(k,l)$ through the coded aperture element $(i,j)$ is given by:
\begin{equation}
(s_x,s_y) = \left( \frac{a_{x(i)}-d_{x(k)}}{L}, \frac{a_{y(j)}-d_{y(l)}}{L} \right) \ ,
\label{eq:sky}
\end{equation}
where $L$ is the distance from the detector plane to the coded aperture plane. Finally, substituting Equation \ref{eq:sky} for Equation \ref{eq:convolution}, we obtain the sky image from the direction $(s_x,s_y)$ by this convolution: 
\begin{equation}
\tilde{S}_{(s_x,s_y)} = \sum_{(i,j)} \sum_{(k,l)} \tilde{A}_{(i,j)} D_{(k,l)} \ .
\end{equation}
\par
For these experiments, we used the synchrotron X-ray beam line BL20B2 \cite{Goto01} in SPring-8 (Super Photon ring - 8 GeV). This is a 215~m-long beam line and the beam can be regarded as parallelly incident X-rays, emulating an infinitely distant celestial source. The 16~keV monochromatic beam came into the coded aperture board and reached the detector. The beam size was 10~mm square, which was sufficiently larger than the size of the patterns 3.5~mm. Since the direction of the beam is fixed, we realized the directional shifts of the X-ray source by leaning the entire detector system with the coded aperture using rotary and goniometer stages (the right side of Figure \ref{setup}) with respect to the beam axis. We generated the observation data of multiple sources (ii) and (iii) by stacking X-ray events at different relative directions. Since this stacking process is done before the decoding calculation, the encoded images of the sources (ii) and (iii) are identical to images which would be obtained by simultaneous observations of the corresponding multiple sources. Due to a very small misalignment between the coded aperture plane and the detector plane, we applied a rotation correction of $15^{\prime \prime}$ in the detection image plane before the decoding process.\par
\begin{figure}[H]
	\centering
	\includegraphics[width = 8 cm]{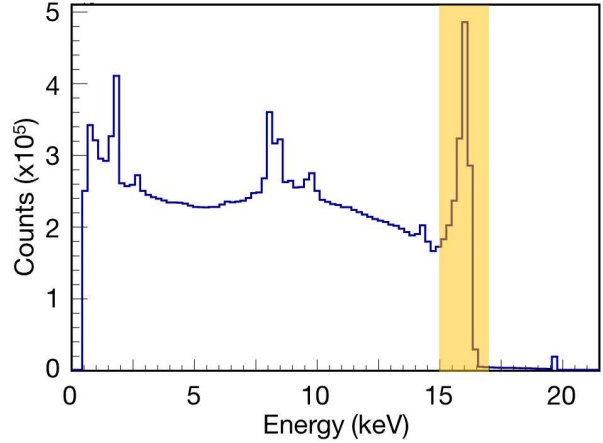}
	\caption[]{An X-ray spectrum for the monochromatic 16~keV beam detected by the CMOS imager. The emphasized area is the energy band from which X-ray events were extracted for the imaging in these experiments.}
	\label{spectrum}
\end{figure}
\begin{figure}[H]
	\centering
	\includegraphics[width = 16 cm]{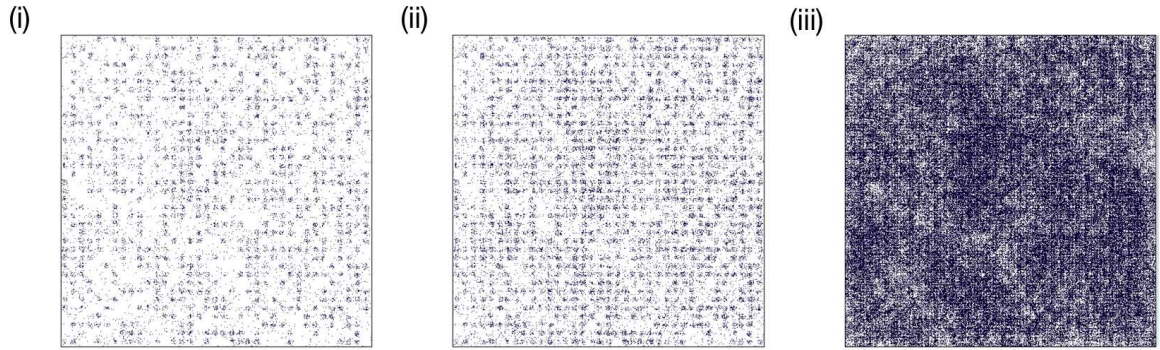}
	\caption[]{The position distributions of X-ray detections in the case of pattern (A) for the datasets in Table \ref{datasets}. These images are shown in linear scales.}
	\label{detected}
\end{figure}

As a detector, we used a 25M pixel CMOS imager with a pixel pitch of 2.5~$\mu$m. This imager has an ability of X-ray detection below 24~keV and has an energy resolution of 176~eV (Full Width at Half Maximum; FWHM) at 5.9~keV at a room temperature of 25~${}^\circ$C \cite{Asakura19}. We set a frame time exposure of 90~ms to reduce detector backgrounds $B_{(k,l)}$ and pedestals in a frame. The pedestals are estimated from the latest set of dark images with the X-ray beam off. To extract X-ray events after subtracting pedestals, we define two kinds of threshold energies. An event threshold is used for determining pixels with X-ray detection. In one X-ray event, the charge generated in a pixel can be extended to its adjacent pixels due to the ejection of a photoelectron and Auger electrons and diffusion of the charge carriers. We also apply a split threshold to the pixels surrounding the X-ray detection pixel in order to take account of pixels with the charge sharing. Pixel values larger than the split threshold are added to the central value to determine the total energy of an event. We set the event and split thresholds to 500 and 50~eV, respectively. Figure \ref{spectrum} shows an X-ray spectrum for the 16~keV beam. We selected events within the 16 keV peak for our imaging experiments as indicated in this spectrum. Then we regard the highest pixel position in an X-ray event as the photon detected position. Figure \ref{detected} shows a distribution of the event positions for each dataset by the pattern (A). We can see a clear shadow of the aperture pattern, though there are some events detected in the area eventually corresponding to the mask or the gaps due to the transmission of the X-ray beam through the SUS104 board ($\sim$10~\% for 16~keV).

\section{Results}
\label{sec:results}
\begin{figure}[H]
	\centering
	\includegraphics[width = 16 cm]{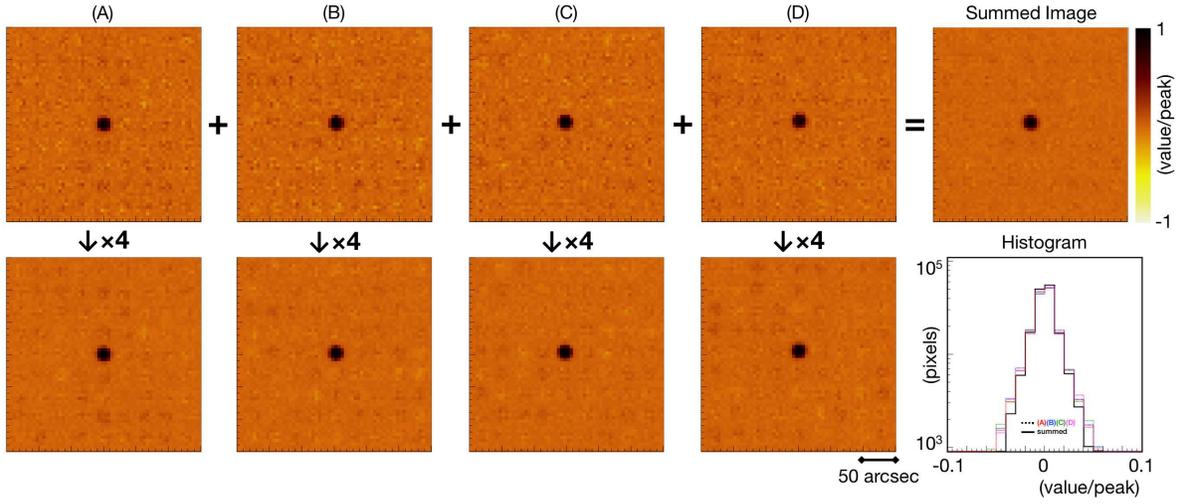}
	\caption[]{Imaging results for the dataset (i) in Table \ref{datasets}. {\it The top-left 4 panels}: The decoded image by each random pattern in Figure \ref{patterns}. These images are binned in 5 arcsec and expanded near the source. The scale is normalized such that the peak value of $\tilde{S}_{(s_x,s_y)}$ is $1$ and the color at 0 is common among all images. {\it The top-right panel}: The summed image of all decoded images. The color scale is made in the same way. {\it The bottom-left 4 panels}: The decoded images with 4 times longer exposure experiments in each pattern. Then the photon statistics is comparable to that of the summed image. The color scale is made in the same way. {\it The lower-right panel}: The histogram of $\tilde{S}_{(s_x,s_y)}$ within the entire FoV excluding the source region considering the angular resolution $\sigma$, i.e. a histogram of the artifact levels. The value 0 means artifact-less. The solid line is for the summed image and colored dashed lines for each long-exposure decoded image, where the color red, blue, green, and magenta represent the pattern (A), (B), (C), and (D) respectively.}
	\label{point}
\end{figure}
\begin{figure}[H]
	\centering
	\includegraphics[width = 16 cm]{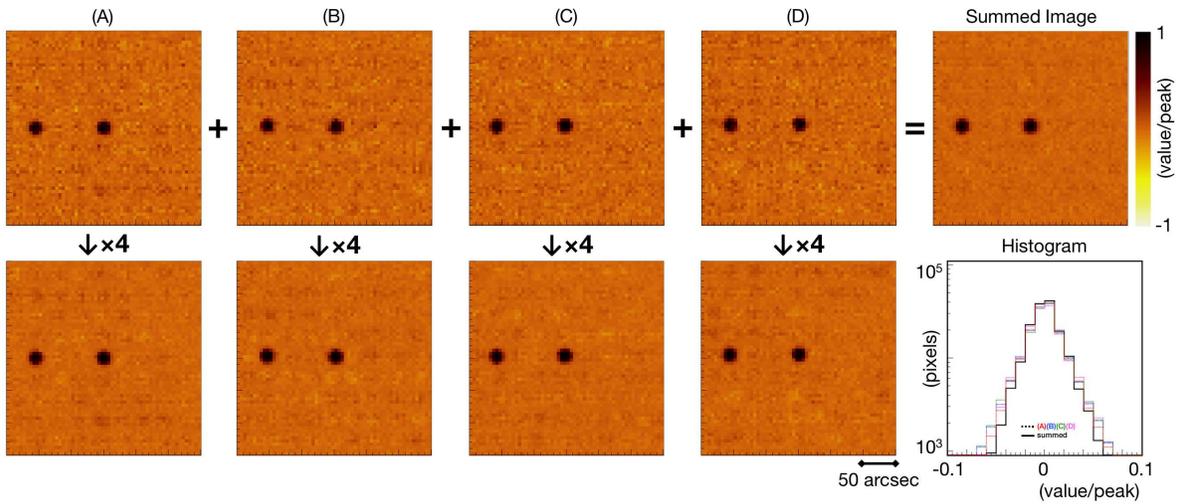}
	\caption[]{The same as Figure \ref{point} but for the dataset (ii).}
	\label{points}
\end{figure}
\begin{figure}[H]
	\centering
	\includegraphics[width = 16 cm]{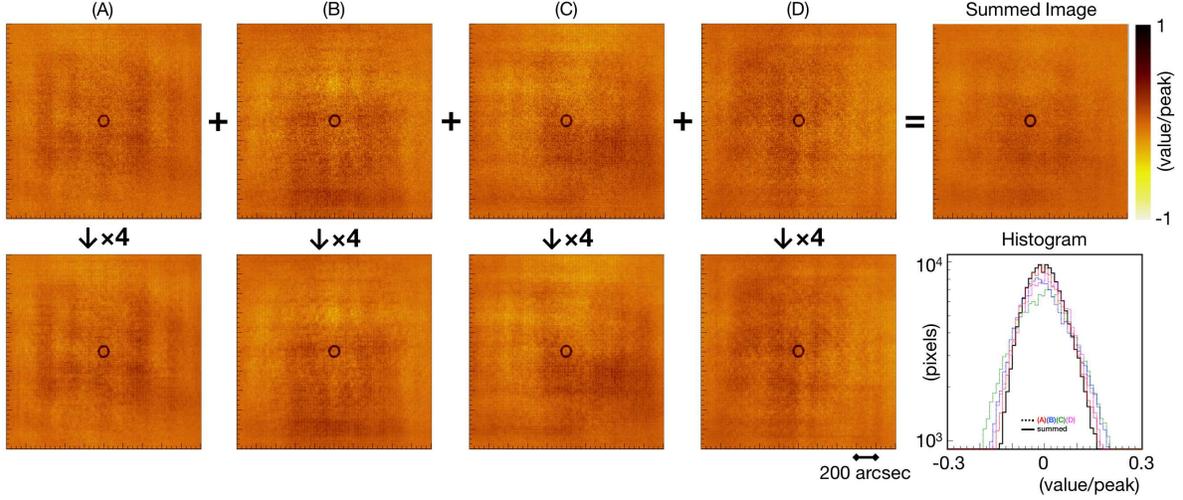}
	\caption[]{The same as Figure \ref{point} but for the dataset (iii). Note that these images are displayed in a wider sky area compared to Figures \ref{point} \& \ref{points} in order to show the artifact pattern distributed in the entire FoV.}
	\label{torus}
\end{figure}
Figures \ref{point}--\ref{torus} show the results of the decoding process for the three datasets in Table \ref{datasets}. The top-left 4 panels of each figure shows the decoded images by all patterns shown in Figure \ref{patterns}. Because these decoded values depend on the decoding process, we show only the ratio to the peak value here in order to compare the magnitude of artifacts. The top-right panel shows the sum of the four decoded images, which is the result of our newly proposed method of the artifact reduction. To compare the effect of the artifact reduction under the same statistical conditions, we demonstrate the situation that we configured 4 same patterns and did the same experiments. The bottom-left 4 panels are the results with 4 times longer exposure time by the same setup. The bottom-right panel shows histograms of the normalized pixel values $\tilde{S}_{(s_x,s_y)}$ of the source-free region in the decoded images to evaluate the effect of summing for multiple different patterns. These histograms give the degree of the fluctuations due to the artifacts.\par
First, we evaluated the performance for point-like sources using dataset (i) and (ii). For the dataset (ii), the detected images are separated by 50~pixels from each other as shown in Figure \ref{detected} (ii). It corresponds to a source separation of $103^{\prime \prime}$. Considering the reading error of 2~pixels and the angular resolution $\sigma$, this is consistent with our decoded image in Figure \ref{points}. This confirms that our decoding method works precisely. As we clearly see in Figures \ref{point} and \ref{points}, the shape of the decoded sources are extended although the actual beam is regarded not to be diffused in the angular space. This extension is due to the angular resolution $\sigma$, and our results imply $\sigma$ (FWHM) is $\sim20^{\prime \prime}$. It is consistent with the calculation that:
\begin{equation}
\sigma = \frac{\sqrt{r^2 + p_{\mathrm{D}}^2}}{L} \ ,
\end{equation}
where $r$ is the diameter of the aperture, $p_{\mathrm{D}}$ is the pixel pitch of the detector, and $L$ is the distance between the coded aperture and the detector. This angular resolution is comparable to that of the {\it NuSTAR} satellite in terms of FWHM $18^{\prime \prime}$ \cite{Harrison13}. This means that our experimental setup achieved an excellent imaging performance with a dramatically downsized imaging system compared to existing missions. Next, the dataset (iii) is for a demonstration of a diffuse source. The imaging system measured a circular source with a radius of $60^{\prime \prime}$. As shown in Figure \ref{torus}, a circle at the center of the FoV is clearly decoded with the precise size considering the angular resolution. Its shape is slightly distorted, which is due to the discrete stage control. 
\begin{table}[ht]
\caption{The standard deviation of normalized $\tilde{S}_{(s_x,s_y)}$ of each histogram in Figures \ref{point}--\ref{torus}.} 
\label{standard_deviation}
\begin{center}       
\begin{tabular}{|c|c|c|c|c|c|} 
\hline
\rule[-1ex]{0pt}{3.5ex} Dataset & (A) & (B) & (C) & (D) & Summed\\
\hline\hline
\rule[-1ex]{0pt}{3.5ex}  (i) & 0.0166 & 0.0173 & 0.0175 & 0.0172 & 0.0143\\
\hline
\rule[-1ex]{0pt}{3.5ex}  (ii) & 0.0236 & 0.0250 & 0.0249 & 0.0247 & 0.0197 \\
\hline
\rule[-1ex]{0pt}{3.5ex}  (iii) & 0.0774 & 0.0895 & 0.1002 & 0.0837 & 0.0704 \\
\hline
\end{tabular}
\end{center}
\end{table} \par
The decoded images in Figures \ref{point}--\ref{torus} show less artifacts in the summed image. For estimating this effect precisely, Table \ref{standard_deviation} shows the standard deviation of each artifact level histogram in Figures \ref{point}--\ref{torus}. This value should be close to 0 in an artifact-free image. Each pattern shows different characteristics. Pattern (A) shows the smallest standard deviation in all datasets. Pattern (C) looks generally the worst, but (B) is slightly worse for the dataset (ii). Pattern (D) shows a similarity to (B) but is good for the dataset (iii). In all the datasets, the summation of the decoded images with the different random patterns reduces the fluctuations of the source-free regions at least by 10 \%. This also quantitatively supports our proposed method of the artifact reduction. 

\section{Discussions}
\label{sec:discuss}
Figures \ref{point}--\ref{torus} and Table \ref{standard_deviation} show the effective reduction of imaging artifacts by using multiple different random patterns. However, the effective area per one coded aperture unit of our experimental setup is very small for astronomical observations in fact. We have to configure parallelly much more than 4 units to gain the total area because the unit size is determined by the scientific requirements. Our method performs more effectively in such a situation by using a different random pattern in each unit.
\begin{figure}[H]
	\centering
	\includegraphics[width = 16 cm]{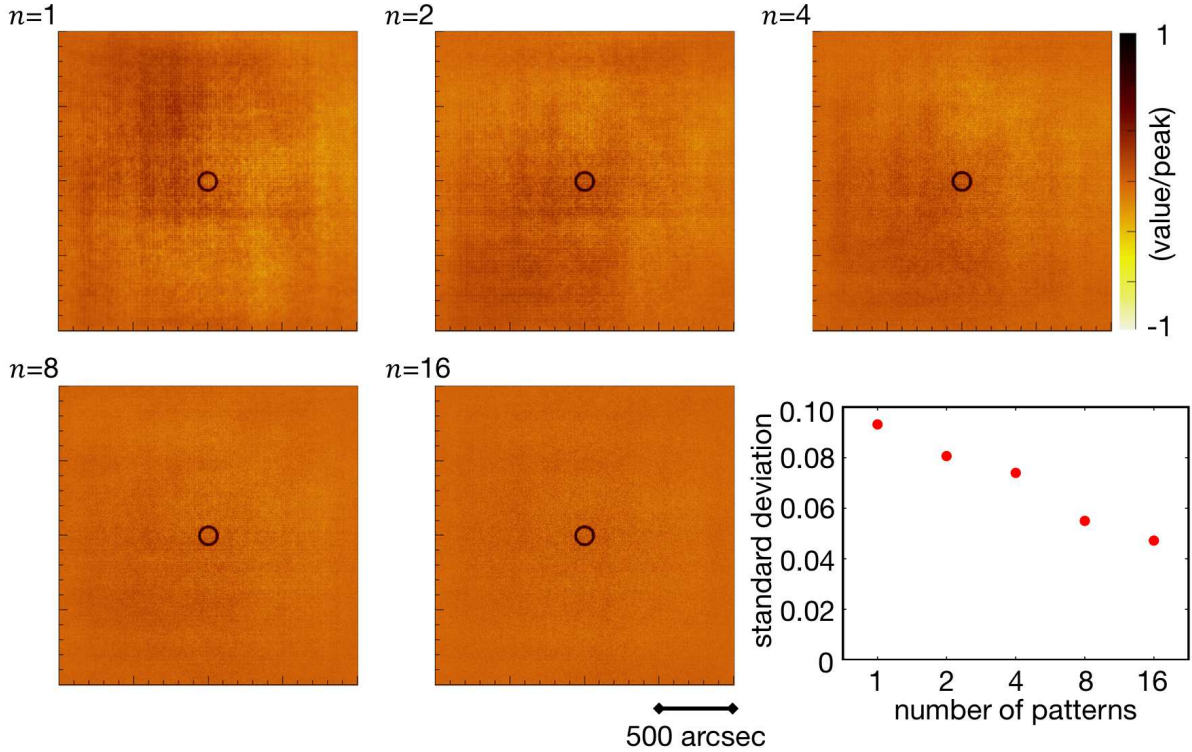}
	\caption[]{{\it Decoded images}: Simulation results. $n$ is the number of patterns we used. {\it The lower right panel}: The standard deviation of $\tilde{S}_{(s_x, s_y)}$ in the source-free regions as a function of the number of patterns.}
	\label{geant4}
\end{figure}\par
We examine the effect of increasing the number of different patterns. It would cost a lot to make this experimentally, so we used Monte Carlo simulations instead here. We used Compton Soft \cite{Odaka10}, which depends on the Geant4 toolkit library for Monte Carlo simulations \cite{Agostinelli03,Allison06,Allison16}. We built the same geometric configuration of our beam experiment (iii), i.e., a 1.5~mm square coded aperture and a detector with a pixel pitch of 2.5~$\mu$m, with a separation of 250~mm. We assumed a circular object with a radius of $60^{\prime \prime}$ on the optical axis, and it emits monochromatic 16 keV photons uniformly inside the thin circle. Concerning the coded apertures, we prepared 16 different random patterns. The number of multiple different random patterns was changed from 1 to 2, 4, 8, and 16, and all simulations were done so that the number of total events kept common to 500,000. Figure \ref{geant4} shows the decoded images together with the standard deviations of all the conditions. In those decoded images, the artifact structure gradually disappears as the number of patterns gets increased, which is the effect consistent with the result of our beam experiment. In particular, artifacts are sufficiently reduced in the 8- and 16- pattern cases. The standard deviation of artifacts also decreases, i.e., the signal-to-noise ratio increases with the number of patterns. Therefore, we are able to gain the effective area and to reduce artifacts simultaneously by using our method. Combining this imaging capability with the polarization detectability of the CMOS imager \cite{Asakura19}, our system can be used for the imaging polarimetry in hard X-rays. \par
For a practical use, we should consider backgrounds as well as celestial signals. We need distinguish two major components: the Cosmic X-ray Background (CXB) and the Non X-ray Background (NXB). The CXB is originated from point sources of active galactic nuclei, distributed almost uniformly in all sky. Our narrow FoV configuration is designed to reduce the contamination of the CXB. It can also restrict the possibility of contaminating bright sources including the appearance of transient sources in the FoV. As to the NXB, which is composed by particles in orbit and secondary X-rays from the satellite and instrument structures, we can estimate it using night Earth occultation and/or a fully masked unit. Considering these practical issues about the effective area and the backgrounds, we stress that our compact system will be promising as a CubeSat mission that targets small bright objects such as solar flares, the Crab nebula, and bright X-ray binaries.\par
Finally, we should note three points about our experiments. First, we can also use a forward fitting method taking account of the projection pattern for the decoding process. This method should work well with appropriate treatment of the statistical model of the signal and background observations, but the problem of systematic artifacts originated from a coded pattern will remain. In addition, It requires more computing costs and then an online processing would not be easy for a limited resource mission. Second, there are other evaluation factors of the imaging performance improved by our method though we only show the artifact intensity spectra. In practice, it would be important to evaluate the point-like source sensitivity and the resolving power of double point sources under the effects of artifacts. These are interesting performance indices but such further quantitative evaluations will be done in the future using a more practical system as an observatory. Third, in our experimental setup the photon positions were highly over-sampled on the detector plane since the pixel pitch of 2.5~$\mu$m is significantly smaller than the requirement by the sampling theorem for the aperture size and pitch. Such over-sampling generally improves the accuracy of imaging and reduces image noises. But the reduction of artifacts shown in Section 4 is primarily due to the multiple random patterns as we propose. These three points do not affect the demonstration of out proposed concept.

\section{Conclusions}
\label{sec:concl}
We propose a new imaging method of the coded aperture using multiple different random patterns simultaneously to decrease systematic artifacts. We actually made an experimental system with 4 different random patterns. Using a 16~keV monochromatic X-ray beam, we achieved a fine angular resolution of $< 30^{\prime \prime}$ (FWHM) with a small configuration of 25~cm. For both point-like and diffuse sources, artifacts decreased effectively in the summed image compared to a single decoded image by each pattern. By Monte Carlo simulations, our concept is shown to be more effective when the number of patterns gets larger.

\acknowledgments 
We thank the anonymous referees for comments to improve the manuscript. We also thank Atsushi Togo for his master thesis, Kentaro Uesugi, Masato Hoshino, Togo Shimozawa, Shigemi Otsuka, Shunsaku Nagasawa, Kairi Mine, Tomoshi Takeda, Yuto Yoshida, and Keisuke Uchiyama for helping our beam experiments, and Yuuki Wada, Hiromasa Suzuki, Tadayuki Takahashi, Noriyuki Narukage, and Kiyoshi Hayashida for helpful suggestions. T.K. is supported by the Advanced Leading Graduate Course for Photon Science (ALPS) in the University of Tokyo. This work is partly supported by the Japan Society for the Promotion of Science (JSPS) KAKENHI Grant Number 19H01906.


\bibliography{report}   
\bibliographystyle{spiejour}   



%

\end{spacing}
\end{document}